\documentclass{iopart}

\usepackage{graphicx}

\expandafter\let\csname equation*\endcsname\relax
\expandafter\let\csname endequation*\endcsname\relax

\usepackage{amsmath}
\usepackage{amsfonts}
\usepackage[dvips]{epsfig}
\usepackage{bm}

\usepackage[T1]{fontenc}
\usepackage[latin9]{inputenc}
\usepackage{float}
\usepackage{amsmath}
\usepackage{graphicx}
\usepackage{amssymb}
\usepackage{bbold}


\begin{document}

\title{Process Tomography for Systems in a Thermal State}

\date{\today}

\author{Ran Ber, Shmuel Marcovitch, Oded Kenneth and\\ Benni Reznik}
\address{School of Physics and Astronomy, Raymond and Beverly Sackler
Faculty of Exact Sciences, Tel-Aviv University, Tel-Aviv 69978, Israel.}

\begin{abstract}
We propose a new method for implementing process tomography which is based on the information extracted from temporal correlations between observables, rather than on state preparation and state tomography.
As such, the approach is applicable to systems that are in mixed states, and in particular thermal states.
We illustrate the method for an arbitrary evolution described by Kraus operators,
as well as for simpler cases such as a general Gaussian channels, and qubit dynamics.

\end{abstract}

\maketitle

\section{Introduction}
Quantum process tomography deals with estimating the dynamics of
unknown systems. It has long been studied due to its fundamental importance
in the fields of quantum communication and quantum computation.

The common approach for implementing quantum process tomography
is based on applying the dynamics on each element of a complete set
of input states, and then performing tomographic measurement of the
output states. This procedure allows one to completely reconstruct the superoperator
representing the dynamics \cite{Poyatos1997}. In recent years several
improvements have been proposed  for this method \cite{Emerson2007,Bendersky2008,Schmiegelow2011},
which reduce its complexity in some cases.

Another approach is based on applying the dynamics on random states
and comparing the results with the theoretical output state which would
have been received had the transformation been purely unitary \cite{Emerson2005}.
The major advantage of this method is that it is efficient (in the sense
that it scales polynomially with the number of particles in the system).
However, it does not reconstruct the superoperator representing the
channel, but rather estimates the strength of the noise.

In the present work we propose a different method for performing process tomography of an unknown evolution,
which is based on the  information extracted from temporal correlations between observables,
and as such does not require state preparation. Therefore, it would be useful in situations were one is restricted to a small set of physical input states and so standard preparation-evolution-measurement scheme cannot be applied. In fact, it can be used with unknown mixed states, and in particular with thermal states. Therefore it is potentially useful in "hot" systems whose state cannot be controlled with current technology, or cooled to the ground state.

The proposed method is based on two key points:
(a) we utilize a proper set of temporal correlations between observables at two consecutive instances of time, say $t_2$ and $t_1$, $t_2>t_1$, that encodes the dynamical evolution of the system, and (b) we employ weak measurements,  rather than ordinary disturbing measurements, in order to measure such correlations. We cannot use ordinary measurements to observe the above temporal correlations because the interaction with the system at the earlier time
will disrupt the correlations. Nevertheless, as we show, by using weak measurements which barely affect the system, such correlations can be measured.

Our method is applicable for general discrete systems whose evolution is described
by a set of Kraus operators, as well as continuous systems whose evolution is linear with respect to a given set of operators.
The rest of the paper is organized as follows:
In the next section we define the temporal covariance matrix which will be sufficient for determining the above type of general evolution. We then show how the temporal covariance can be measured using weak measurements.
In section 3 we show how to find
the Kraus operators for general discrete systems, in section 4 we show how to find the dynamics for Gaussian channels, in section 5 we demonstrate the proposed method on qubit dynamics and in section 6 we present a numerical study.

\section{Measurement of a temporal covariance matrix}

Let us begin by defining the two-point temporal covariance matrix:
\begin{eqnarray}
\sigma_{ij}\!\left(t_{1},t_{2}\right)\!&\!\equiv & \left\langle \left\{B_{i}\!\left(t_{1}\right)\!-\!\left\langle B_{i}\!\left(t_{1}\right)\right\rangle \! ,\!B_{j}\!\left(t_{2}\right)\!-\!\left\langle B_{j}\left(t_{2}\right)\right\rangle \right\} \right\rangle \nonumber \\
 & = & 2{\rm Re}\!\left[\left\langle B_{i}\!\left(t_{1}\right)B_{j}\!\left(t_{2}\right)\right\rangle -\left\langle B_{i}\!\left(t_{1}\right)\right\rangle \left\langle B_{j}\!\left(t_{2}\right)\right\rangle \right],\nonumber \\
\label{definition of the covariance matrix}
\end{eqnarray}
where $\left\{B_{i}\right\}$ are Hermitian operators. This unusual definition of the covariance matrix, where the measurements
are taken at different times, proves to be essential for our method.

In order to measure the covariance matrix one has to measure the observables $B_{i}\left(t\right)$ and the correlations $\left\{ B_{i}\left(t_{1}\right),B_{j}\left(t_{2}\right)\right\}$. While the measurement of the observables is straightforward, the measurement of the correlations is not trivial and the rest of this section is devoted to it.

The general scheme for the measurement of a single correlation $\left\{ B_{i}\left(t_{1}\right),B_{j}\left(t_{2}\right)\right\}$
involves two measurement devices: one measures $B_{i}$ at time $t_1$, and the other measures $B_{j}$ at time $t_2$. In order to measure the correlation one has to measure a joint operator of both measurement devices.
Our method can be implemented in numerous ways. For simplicity, we shall
demonstrate it with spin pointers and assume unitary time evolution.
In this case, the Hamiltonian of the coupling between the system and
the $n$'th pointer measuring $B_{i}$ is $H_{ni}=\frac{\epsilon}{2}B_{i}\otimes\sigma_{y}^{[n]}\delta\left(t\right)$
(where $\epsilon\ll1$), therefore the measurement propagator is
\begin{equation}
U_{ni} = e^{-i\frac{\epsilon}{2}B_{i}\otimes\sigma_{y}^{[n]}} = \cos\left(\frac{\epsilon}{2}B_{i}\right)-i\sin\left(\frac{\epsilon}{2}B_{i}\right)\otimes\sigma_{y}^{[n]}\,.
\end{equation}
Assuming both pointers initially point at $\left|\uparrow_{x}\right\rangle \equiv\frac{1}{\sqrt{2}}\left(\left|\downarrow_{z}\right\rangle +\left|\uparrow_{z}\right\rangle \right)$,
we denote the initial state as $\rho_{\text{in}}\equiv \underset{\text{a}}{\overset{}{\sum}}p_{\text{a}}\left|\psi_{\text{a,sys}},\uparrow_{x},\uparrow_{x}\right\rangle \left\langle\psi_{\text{a,sys}},\uparrow_{x},\uparrow_{x}\right|$, where the label 'sys' stands for 'system'.
Setting $\rho_{\text{in}}=\rho\left(t_{1}\right)$, the final state, $\rho_{\text{fi}}$, is given by
\begin{eqnarray}
\rho_{\text{fi}}  & = & U\left(t_{2}\right)U^{\dagger}\left(t_{2}\right)U_{2j}U\left(t_{2}\right)U_{1i}\rho_{\text{in}}U_{1i}^{\dagger}U^{\dagger}\left(t_{2}\right) U_{2j}^{\dagger}U\left(t_{2}\right)U^{\dagger}\left(t_{2}\right) \nonumber \\
 & = & U\left(t_{2}\right)\left(\cos\left(\frac{\epsilon}{2}B_{j}\left(t_{2}\right)\right)-i\sin\left(\frac{\epsilon}{2}B_{j}\left(t_{2}\right)\right)\otimes\sigma_{y}^{[2]}\right)\nonumber \\
 &  & \times\left(\cos\left(\frac{\epsilon}{2}B_{i}\left(t_{1}\right)\right)-i\sin\left(\frac{\epsilon}{2}B_{i}\left(t_{1}\right)\right)\otimes\sigma_{y}^{[1]}\right)\rho_{\text{in}}\nonumber \\ & & \times\left(\cos\left(\frac{\epsilon}{2}B_{i}\left(t_{1}\right)\right)+i\sin\left(\frac{\epsilon}{2}B_{i}\left(t_{1}\right)\right)\otimes\sigma_{y}^{[1]}\right)\nonumber \\ & & \times\left(\cos\left(\frac{\epsilon}{2}B_{j}\left(t_{2}\right)\right)+i\sin\left(\frac{\epsilon}{2}B_{j}\left(t_{2}\right)\right)\otimes\sigma_{y}^{[2]}\right)U^{\dagger}\left(t_{2}\right) ,\label{rho final}
\end{eqnarray}
where $U\left(t_{2}\right)$ is the time propagator.
The pointers' expectation value and variance are therefore
\begin{eqnarray}
E\!\left(\sigma_{z}^{[1]}\!\otimes\!\sigma_{z}^{[2]}\right)\!& \!=\! &\!\frac{\epsilon^{2}}{2}\text{Tr}\!\left(\left\{ B_{i}\left(t_{1}\right)\!,\!B_{j}\left(t_{2}\right)\right\} \!\rho_{\text{in}}\right)\!+\!\epsilon^{4}f\!\left(B_{i}\left(t_{1}\right)\!,\!B_{j}\left(t_{2}\right)\right)\!+\!\mathcal{O}\!\left(\!\epsilon^{6}\!\right)\!
,\label{pointers expectation value}\\
V\!\left(\sigma_{z}^{[1]}\!\otimes\!\sigma_{z}^{[2]}\right) & = & 1+\mathcal{O}\left(\epsilon^{4}\right).\label{pointers variance}
\end{eqnarray}
where $\left|f\left(B_{i}\left(t_{1}\right),B_{j}\left(t_{2}\right)\right)\right|\leq\frac{1}{3}\left\Vert B\right\Vert ^{4}$ (the notion $\left\|\cdot\right\|$ stands for operator norm \cite{OperatorNorm}), for more details see appendix III. This procedure can be easily generalized to non-unitary time evolutions and pointers of general dimension. Similar calculations for continuous pointers are found in \cite{Mitchison2007}. Note that while the error described in Eq. (\ref{pointers variance}) is random, the error described in Eq. (\ref{pointers expectation value}) is systematic. This means that in principle, for a given system, one can calculate $f\left(B_{i}\left(t_{1}\right),B_{j}\left(t_{2}\right)\right)$ (either exactly or perturbatively), fix the systematic error and thus dramatically improve the efficiency of this process. This direction will be discussed further in section 6.

It can be shown that following a single correlation measurement procedure
(for short times),
\begin{equation}
\rho_{{\rm fi,sys}}=\rho_{{\rm in,sys}}+\mathcal{O}\left(\epsilon^{2}\right)\,,\label{rho_s_final =00003D00003D  rho_s_initial + O(epsilon3)}
\end{equation}
i.e., the state is hardly influenced by the measurement, hence we refer to this measurement as "weak" \cite{Aharonov1988,Aharonov2005} (Note that we do not use post selection). This is the key feature that enables this measurement method to work. As shown in equations (\ref{pointers expectation value}) and (\ref{pointers variance}) the "weakness" of the measurement comes with a price, every single weak measurement is highly inaccurate. This can be compensated by a large number of measurements.
The error after $N$ such measurements of a single correlation is $\delta=\max\left\{\frac{2}{\epsilon^{2} \sqrt{N}},\left|f\!\left(B_{i}\left(t_{1}\right)\!,\!B_{j}\left(t_{2}\right)\right)\right|\epsilon^{2}\right\}$, hence in order to reach this error the optimal measurement strength is $\epsilon=\sqrt{\delta\left|f\left(B_{i}\left(t_{1}\right),B_{j}\left(t_{2}\right)\right)\right|^{-1}}$ and the number of measurements required is $N=\frac{4f^{2}\left(B_{i}\left(t_{1}\right),B_{j}\left(t_{2}\right)\right)}{\delta^{4}}\leq\frac{4}{9}\frac{\left\Vert B\right\Vert ^{8}}{\delta^{4}}$. Since this estimate was derived for $\epsilon\ll 1$ it is valid only for $\delta\ll\frac{1}{3}\left\|B\right\|^{4}$.

In appendix II we present an alternative method for measuring the correlations. This alternative method involves only single pointer measurements, and so we believe this new method may be easier to implement.

\section{Constructing the dynamics for discrete systems}

In this section we show how the dynamics can be estimated using the two-point temporal covariance matrix.
Given a $D$ level system, we wish to define
a complete basis of operators $\left\{ B_{a}\right\}\equiv\left\{ B_{0},B_{i}\right\}$
($i\in\left\{ 1..D^{2}-1\right\} $), such that $B_{0}=\mathbb{1}_{D\times D}$ and $\left\{B_{i}\right\}$
are chosen to be Hermitian matrices.
In the Heisenberg picture, since the set of operators is complete, and superoperators representing quantum channels are linear, the following equation holds
\begin{equation}
B_{i}\left(t\right)=M_{ij}\left(t,t_{0}\right)B_{j}\left(t_{0}\right)+\chi_{i}\left(t,t_{0}\right)\,.\label{Bi(t) as a function of B(j) and B(0) (short)}
\end{equation}
Substituting this equation in Eq. (\ref{definition of the covariance matrix}) we obtain
\begin{equation}
\sigma_{ik}\left(t,t_{0}\right)=M_{ij}\left(t,t_{0}\right)\sigma_{jk}\left(t_{0},t_{0}\right)\,.\label{sigma(t,t0) as a function of sigma(t0,t0)}
\end{equation}
This equation defines the time evolution of the covariance matrix. $M$ could be retrieved from the covariance matrices simply by
\begin{equation}
M\!\left(t,t_{0}\right)=\sigma\left(t,t_{0}\right)\sigma^{-1}\left(t_{0},t_{0}\right)\,. \label{M as a function of Sigma}
\end{equation}
In general, $\sigma\left(t_{0},t_{0}\right)$ might be singular. This happens if and only if the initial density matrix is singular (see appendix I for a proof). In this case the density operator doesn't sample all the states in the Hilbert space, and so it is impossible to gain a complete knowledge of the evolution. However, it is worth mentioning that when picking a random initial state over a continuous distribution, the chance of producing a singular density matrix is zero.

From Eq. (\ref{Bi(t) as a function of B(j) and B(0) (short)}) we get
\begin{equation}
\chi\left(t,t_0\right)=\left\langle B\left(t\right)\right\rangle -\sigma\left(t,t_{0}\right)\sigma^{-1}\left(t_{0},t_{0}\right)\left\langle B\left(t_{0}\right)\right\rangle \,.\label{chi as a function of sigma}
\end{equation}
Note that the estimation of $\chi\left(t,t_0\right)$ does not require measurements in addition to the ones needed in order to determine $\sigma\left(t,t_0\right)$. Since $M$ and $\chi$ entirely encode the dynamics, one can fully estimate the dynamics by measuring the two-point temporal covariance matrix.

Since this method relies on the measurement of $\sigma_{ij}\left(t_0,t_0\right)$ and $\sigma_{ij}\left(t,t_0\right)$ for every $i,j\in\left\{ 1..D^{2}-1\right\}$ its complexity grows with $D$ as $\mathcal{O}\left(D^4\right)$. For $n$ qudits ($d$ level systems) $D=d^{n}$ and thus the complexity grows with the number of qudits as $\mathcal{O}\left(d^{4n}\right)$, the same as in \cite{Poyatos1997}.

Note that the described method works for general non-singular states, and in particular maximally mixed states and thermal states. An alternative procedure, based on covariance matrices defined at one time only, would not work with general states.

While $M$ and $\chi$ entirely encode the dynamics, it would be convenient to describe the channel using the Kraus representation.
The Kraus representation of a superoperator is
\begin{equation}
\rho\left(t\right)=\underset{\mu}{\overset{}{\sum}}K_{\mu}\left(t,t_{0}\right)\rho\left(t_{0}\right)K_{\mu}^{\dagger}\left(t,t_{0}\right)\,,\label{Kraus equation}
\end{equation}
where the following constraint holds
\begin{equation}
\underset{\mu}{\overset{}{\sum}}K_{\mu}^{\dagger}\left(t,t_{0}\right)K_{\mu}\left(t,t_{0}\right)=\mathbb{1}.\label{Sigma Kdagger K =00003D00003D I}
\end{equation}
By transferring the time dependence from the Schrodinger picture to
the Heisenberg picture, demanding that the expectation value of a
general operator $\hat{O}$ remain unaffected by the picture, one
can deduce the equivalent equation for operators:
\begin{equation}
\hat{O}\left(t\right)=\underset{\mu}{\overset{}{\sum}}K_{\mu}^{\dagger}\left(t,t_{0}\right)\hat{O}\left(t_0\right)K_{\mu}\left(t,t_{0}\right)\,.\label{Kraus equation for operators}
\end{equation}
Since the set of operators $\left\{ B_{a}\right\}\equiv\left\{ B_{0},B_{i}\right\}$ ($B_{a}\equiv B_{a}\left(t_0\right)$) is complete,
the Kraus operators can be represented as
\begin{equation}
K_{\mu}\left(t,t_{0}\right)=u_{a\mu}\left(t,t_{0}\right)B_{a}\label{Kraus operators constructed from basis operators}
\end{equation}
(summation on double indices is assumed) where $u_{\mu}$ are complex vectors of coefficients and the following relations
hold:
\begin{eqnarray}
\left[B_{a},B_{b}\right] & = & if_{abc}B_{c}\,,\label{Definition of f}\\
\left\{ B_{a},B_{b}\right\}  & = & g_{abc}B_{c}\,,\label{Definition of d}
\end{eqnarray}
where $f_{abc}$ and $g_{abc}$ determine the structure of the basis.
Substituting Eq. (\ref{Kraus operators constructed from basis operators})-(\ref{Definition of d})
in Eq. (\ref{Kraus equation for operators}) we obtain:
\begin{equation}
B_{a}\left(t\right)=\frac{1}{4}u_{b\mu}^{*}u_{c\mu}\left(if_{bad}+g_{bad}\right)\left(if_{dce}+g_{dce}\right)B_{e}\left(t_0\right)\,,\label{Ba(t) as a function of Be(0)}
\end{equation}
In order to bring this equation to the form of Eq. (\ref{Bi(t) as a function of B(j) and B(0) (short)}) we set $M_{ij}\left(t,t_{0}\right)\equiv\frac{1}{4}u_{b\mu}^{*}u_{c\mu}\left(if_{bid}+g_{bid}\right)\left(if_{dcj}+g_{dcj}\right)$
and $\chi_{i}\left(t,t_{0}\right)\equiv u_{b\mu}^{*}u_{c\mu}\left(if_{bid}+g_{bid}\right)\left(if_{dc0}+g_{dc0}\right)$.
Next, we use additional constraints on $u_{a\mu}$'s which are derived by
substituting Eq. (\ref{Kraus operators constructed from basis operators})
in Eq. (\ref{Sigma Kdagger K =00003D00003D I})
\begin{equation}
\frac{1}{2}u_{a\mu}^{*}u_{b\mu}\left(if_{abc}+g_{abc}\right)B_{c}=\mathbb{1}\,,
\end{equation}
which means
\begin{eqnarray}
\frac{1}{2}u_{a\mu}^{*}u_{b\mu}\left(if_{abk}+g_{abk}\right) & = & 0\label{Constraint 1}\,, \\
\frac{1}{2}u_{a\mu}^{*}u_{b\mu}\left(if_{ab0}+g_{ab0}\right) & = & 1\label{Constraint 2}\,.
\end{eqnarray}
The symmetrical part of $M$, anti-symmetrical
part of $M$ and the trace of $M$, together with $\chi$ and the
two constraints of Eq. (\ref{Constraint 1}) and (\ref{Constraint 2})
form 6 equations for 6 unknown objects which are $u_{0\mu}^{*}u_{0\mu}$,
$\text{Re}\left(u_{i\mu}^{*}u_{0\mu}\right)$, $\text{Im}\left(u_{i\mu}^{*}u_{0\mu}\right)$,
$\text{Re}\left(u_{i\mu}^{*}u_{j\mu}\right)$, $\text{Im}\left(u_{i\mu}^{*}u_{j\mu}\right)$
and $u_{i\mu}^{*}u_{i\mu}$. The solution of this set of equations
fully determines the matrix $u_{\mu}\otimes u^{\dagger}_{\mu}$. Since this matrix is
Hermitian it can be diagonalized. Using the resulting sets of
$D$ orthonormal eigenvectors $\left\{ v_{\mu}\right\} $ and corresponding
eigenvalues $\left\{ \lambda_{\mu}\right\} $, we obtain $u_{\mu}=\sqrt{\lambda_{\mu}}v_{\mu}$
(no summation on $\mu$), and thus a possible set of $D^{2}$ Kraus operators that govern
the system's dynamics is given by
\begin{equation}
K_{\mu}=\sqrt{\lambda_{\mu}}v_{\mu a}B_{a}\,.\label{K's as a function of the dynamics for spins}
\end{equation}

Given the ability to measure $M$ and $\chi$ at small time intervals it is possible to estimate the time derivative of the Kraus operators. For Markovian systems this allows one to estimate the Lindblad equation as described in \cite{Lidar2001}.

\section{Constructing the dynamics for Gaussian channels}

In general, our proposed method requires the use of a complete set of operators.
Nevertheless, in special cases it can be applied using a limited set of observables. One important example for this statement is the class of systems described by quadratic Hamiltonians $H=\lambda_{ij}\eta_{i}\eta_{j}+\alpha_{i}\eta_{i}$ ($\eta\equiv\left(\bar{x},\bar{p}\right)$ where $\bar{x}$ and $\bar{p}$ are arrays of coordinate and conjugate momenta respectively). The evolution of a subset of the system is then described as a Gaussian channel \cite{Eisert2005,Holevo2012} that dictates for the conjugate coordinates an evolution of the type
\begin{equation}
\eta_{i}\left(t\right) = M_{ij}\left(t,t_{0}\right)\eta_{j}\left(t_0\right) + \chi_{i}\left(t,t_{0}\right) \label{eta(t) = M eta(0) + chi}\,.
\end{equation}
This equation is identical to Eq. (\ref{Bi(t) as a function of B(j) and B(0) (short)}). Therefore the process of estimating the matrix $M$ is the same as described in the previous section. For Gaussian states $M$ is sufficient in order to calculate the density matrix evolution in time.
While in the case of discrete systems the density matrix must be non singular in order for $\sigma\left(t_0,t_0\right)$ to be non singular, in the case of the Gaussian channels, the method would work for every state, even for states that do not sample the whole basis of states. The proof goes as follows:
according to a variant of the uncertainty principle \cite{Williamson1936,Simon1987} for every covariance
matrix of canonical operators there exists a symplectic matrix
$S$ and a diagonal matrix $W$ where $W_{ii}\geq\frac{1}{2}$ which satisfies $W=S\sigma\left(t,t\right)S^{{\rm T}}$.
Since $S$ is symplectic it is invertible and so $\sigma\left(t,t\right)=S^{-1}W\left(S^{{\rm T}}\right)^{-1}$.
In this form its clear that $\sigma\left(t,t\right)^{-1}=S^{{\rm T}}W^{-1}S$
always exists, and in particular for $t=t_0$. The meaning of this feature is that in the case of Gaussian channels, our method is truly state independent.

Since this method relies on the measurement of $\sigma_{ij}\left(t_0,t_0\right)$ and $\sigma_{ij}\left(t,t_0\right)$ for every $i,j\in\left\{ 1..2n\right\}$ (where $n$ is the number of particles) its complexity grows with the number of particles as $\mathcal{O}\left(n^2\right)$.

It is interesting to remark that in this Gaussian case, there is an analogy between the proposed method
and Gaussian state tomography. In the "spatial problem" it is well known that the spatial correlations encoded in
the covariance matrix fully determine a Gaussian state \cite{Holevo2011}. In our case we see that the Gaussian dynamics is encoded in the temporal correlations which form a temporal correlation matrix.

We believe Gaussian channel could be among the most practical channels to estimate using our method. A simple realization of a single particle Gaussian channel would be an ion inside a trap. The spatial degrees of freedom will be regarded as the system and the spin will be used as the measurement device. Before the beginning of the estimation process the system can conveniently be prepared in a thermal state, while the measurement device has to be in a pure state. The interaction between the system and the measurement device could be implemented using lasers. Following each measurement the spin would have to be brought back to the pure state, however no initialization is required for the system. The only time consuming process in this procedure is the initialization of the spin.

\section{Example: Constructing the Kraus operators for qubit dynamics}

For qubit dynamics we choose the natural operator basis $\left\{ B_{0}=\frac{1}{\sqrt{2}}\mathbb{1},B_{i}=\frac{1}{\sqrt{2}}\sigma_{i}\right\} $.
The basis structure is therefore
\begin{eqnarray}
f_{ijk} & = & \sqrt{2}\epsilon_{ijk}\,,\\
g_{ab0}=g_{a0b}=g_{0ab} & = & \sqrt{2}\delta_{ab}\,,
\end{eqnarray}
and the rest of the coefficients vanish.
Using the explicit form of $f_{abc}$ and $g_{abc}$ we calculate
\begin{eqnarray}
M_{ij} & = & \frac{1}{2}\left(u_{0\mu}^{*}u_{0\mu}-u_{k\mu}^{*}u_{k\mu}\right)\delta_{ij}\nonumber \\
 &  & +\text{Im}\left(u_{0}^{*}u_{k}\right)\epsilon_{ijk}+\text{Re}\left(u_{i}^{*}u_{j}\right)\,,\label{M for spins}\\
\chi_{i}\left(t,t_{0}\right) & = & \text{Re}\left(u_{0}^{*}u_{i}\right)+\frac{1}{2}\text{Im}\left(u_{j}^{*}u_{k}\right)\epsilon_{ijk}\,.\label{chi for spins}
\end{eqnarray}
Recalling that in this case $D=2$ we obtain
\begin{eqnarray}
\text{Tr}\left(M\right) & = & u_{0\mu}^{*}u_{0\mu},\\
M_{ij}\!+\! M_{ji} & = & \!\left(u_{0\mu}^{*}u_{0\mu}\!-\! u_{k\mu}^{*}u_{k\mu}\right)\!\delta_{ij}\!+\!2\text{Re}\!\left(u_{i\mu}^{*}u_{j\mu}\right),\\
M_{ij}\!-\! M_{ji} & = & 2\text{Im}\left(u_{0\mu}^{*}u_{k\mu}\right)\epsilon_{ijk}.
\end{eqnarray}
From the constraints of Eq. (\ref{Constraint 1}) and (\ref{Constraint 2})
we obtain two additional equations:
\begin{eqnarray}
\text{Re}\left(u_{0\mu}^{*}u_{k\mu}\right)-\frac{1}{2}\text{Im}\left(u_{i\mu}^{*}u_{j\mu}\right)\epsilon_{ijk} & = & 0\,,\label{First constraint for spins}\\
u_{0\mu}^{*}u_{0\mu}+u_{i\mu}^{*}u_{i\mu} & = & \sqrt{2}\,.\label{Second constraint for spins}
\end{eqnarray}
The solution of equations (\ref{chi for spins})-(\ref{Second constraint for spins}) is
\begin{eqnarray}
u_{0\mu}^{*}u_{0\mu} & = & \frac{1}{2}\text{Tr}\left(M\right)\,,\label{u0*u0 for spins}\\
\text{Re}\left(u_{i\mu}^{*}u_{0\mu}\right) & = & \frac{1}{2}\chi_{i}\,,\label{Re(ui*u0 for spins)}\\
\text{Im}\left(u_{i\mu}^{*}u_{0\mu}\right) & = & \frac{1}{4}\epsilon_{ijk}\left(M_{kj}-M_{jk}\right)\,,\label{Im(ui*u0 for spins)}\\
\text{Re}\left(u_{i\mu}^{*}u_{j\mu}\right) & = & \frac{1}{2}\left(M_{ij}+M_{ji}\right)+\left(\frac{1}{\sqrt{2}}-\text{Tr}\left(M\right)\right)\delta_{ij}\,,\label{Re(ui*uj for spins)}\\
\text{Im}\left(u_{i\mu}^{*}u_{j\mu}\right) & = & \frac{1}{2}\epsilon_{ijk}\chi_{k}\,.\label{Im(ui*uj for spins)}
\end{eqnarray}
Equations (\ref{u0*u0 for spins})-(\ref{Im(ui*uj for spins)}) construct the matrix
$u_{\mu}\otimes u_{\mu}^{\dagger}$ which is used to calculate a possible set of Kraus operators as explained above. This result can be easily generalized to $n$ interacting qubits.

\begin{figure}[H]
\includegraphics[scale=0.5]{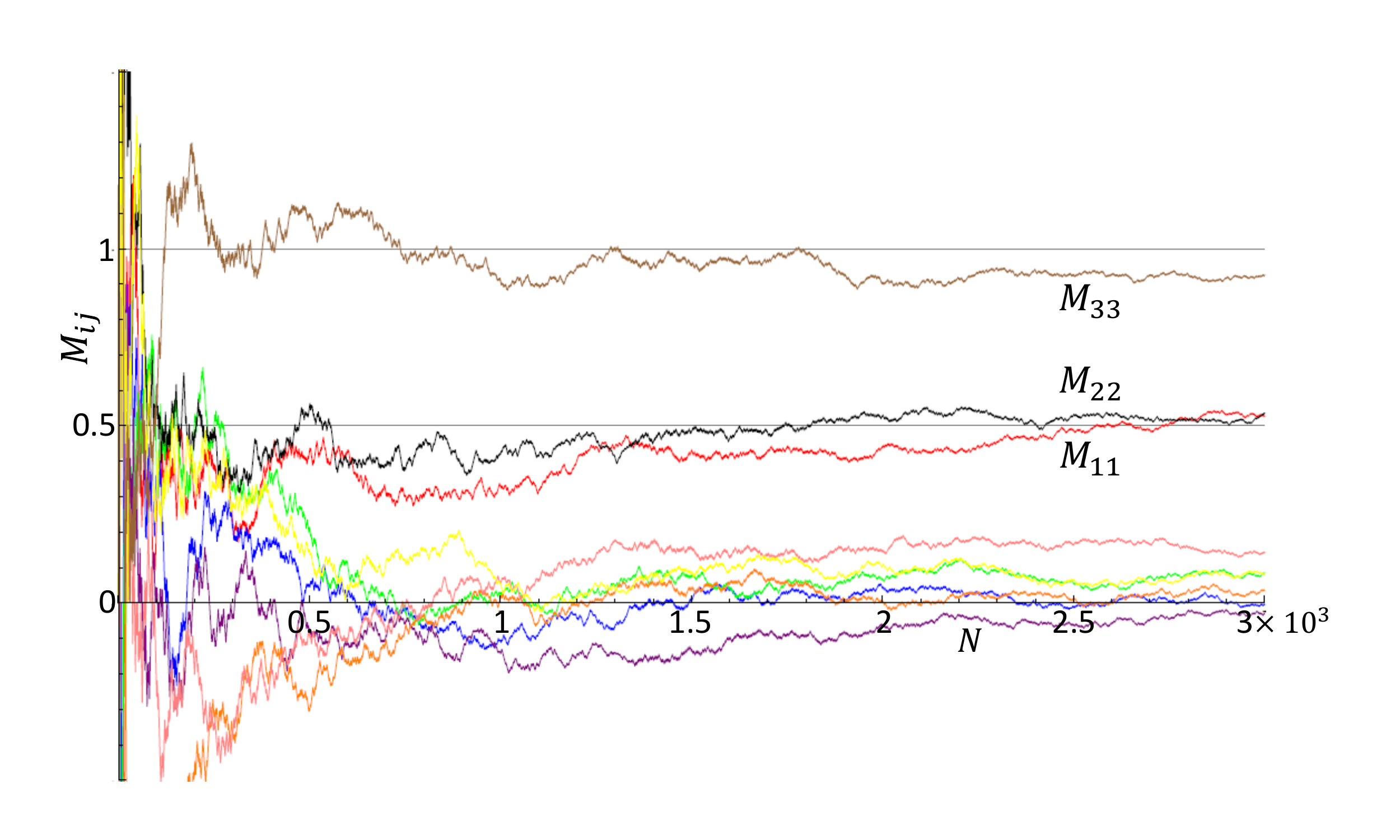}
\caption{Simulations of a phase-damping channel ($p=0.5$) estimation. The set of basis operators was chosen to be $\left\{ B_{i}\right\} =\left\{ \sigma_{i}/\sqrt{2}\right\}$ .Under this setting the theoretical elements of $M$ are $M_{11}=M_{22}=0.5$, $M_{33}=1$ while all the rest are zero. The initial state was chosen to be $\rho_{\text{in}}={1\over 2}\mathbb{1}$. The graph shows the estimated elements of $M$ as a function of the number of measurements per correlation for $\epsilon^{2}=4/9$. Red: $M_{11}$, blue: $M_{12}$, green: $M_{13}$, purple: $M_{21}$, black: $M_{22}$, orange: $M_{23}$, pink: $M_{31}$, yellow: $M_{32}$, brown: $M_{33}$. Note that due to the systematic error the estimated values do not converge exactly to the theoretical values, however an error of $\Delta M_{ij}<0.1$ is reached for $8$ out of $9$ elements of $M$ within $2,500$ measurements per correlation.}
\label{Simulation1}
\end{figure}

\begin{figure}[H]
\includegraphics[scale=0.5]{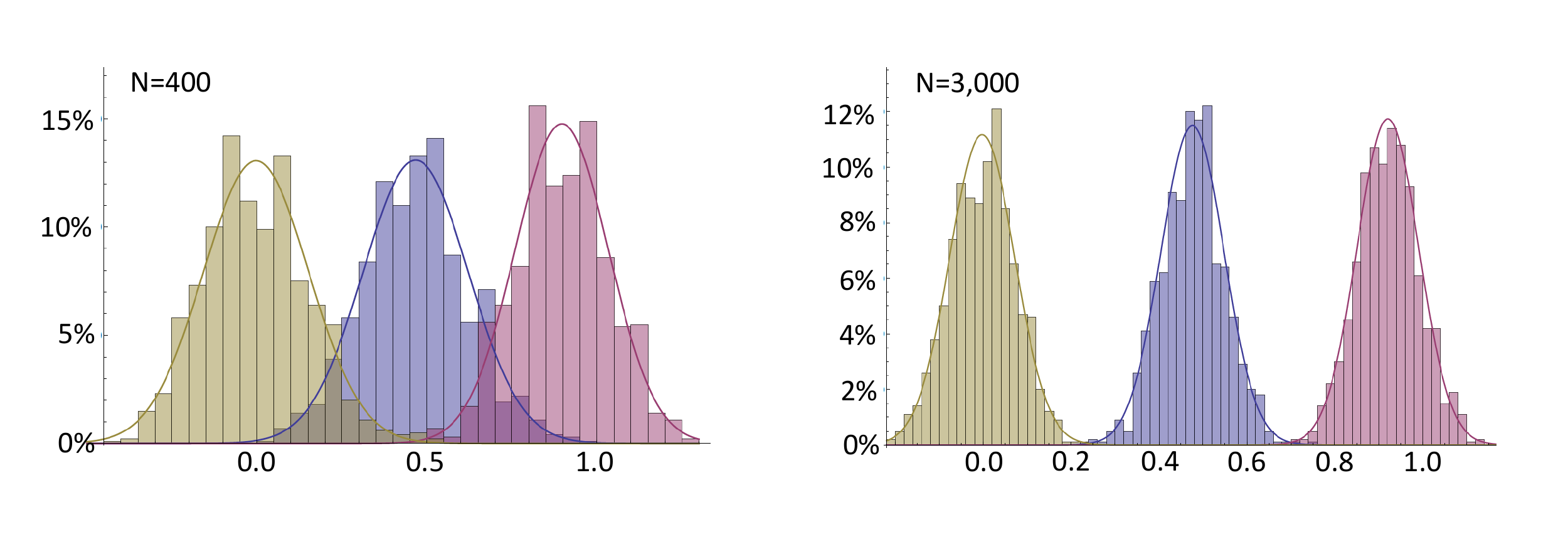}
\caption{Histogram of estimated values of $M_{12}$ (brown), $M_{11}$ (blue) and $M_{33}$ (purple) representing $1,000$ different simulations. Left: $400$ measurements per correlation. The estimated results are $M_{12}=0.00\pm 0.15$, $M_{11}=0.47\pm 0.15$ and $M_{33}=0.90\pm 0.14$. Right: $3,000$ measurements per correlation. The estimated results are $M_{12}=0.00\pm 0.07$, $M_{11}=0.48\pm 0.07$ and $M_{33}=0.92\pm 0.07$. Note that the mean values do not always agree with the theoretical value due to the systematic error.}
\label{Simulation2}
\end{figure}

\section{Numerical study}

\begin{figure}[H]
\includegraphics[scale=0.5]{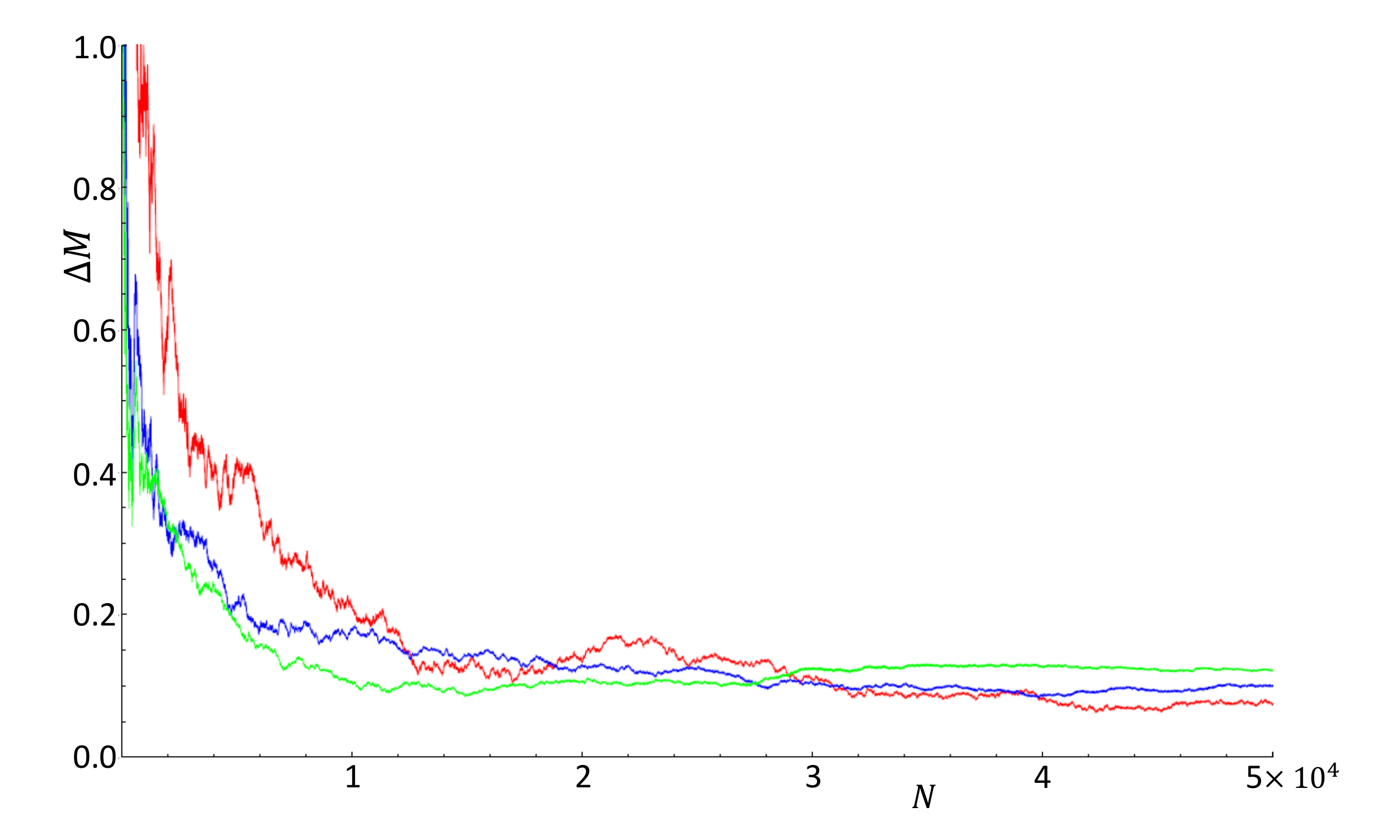}
\caption{$\Delta M$ as a function of the number of measurements per correlation. Red: $\epsilon^{2}=2/9$, blue: $\epsilon^{2}=4/9$, green: $\epsilon^{2}=6/9$. Note that stronger couplings converge to their minimal value faster, but their minimal value is larger due to the larger systematic error. These results exceed the bound $N\leq\frac{4}{9}\frac{\left\Vert B\right\Vert ^{8}}{\delta^{4}}$ since the condition $\delta\ll\frac{1}{3}\left\|B\right\|^{4}$ wasn't satisfied in the settings we chose for the simulation.}
\label{Simulation3}
\end{figure}

We have simulated the channel estimation process for several channels and initial states.
In Figs. (\ref{Simulation1}-\ref{Simulation3}) we show simulations of a phase-damping channel estimation process. Fig. (\ref{Simulation1}) presents a single channel estimation process, Fig. (\ref{Simulation2}) presents histograms of estimated values of $M_{11}$, $M_{12}$ and $M_{33}$ obtained by $1,000$ different simulations and Fig. (\ref{Simulation3}) presents $\Delta M\equiv\left\Vert M_{\text{estimated}}-M_{\text{theoretical}}\right\Vert$ as a function of the number of measurements, $N$, for numerous measurement couplings.

As shown in Section 2, for the method presented here, the number of measurements required to reach an error of $\delta$ for each element of $M$ scales as $\mathcal{O}\left(\delta^{-4}\right)$. For the standard method \cite{Poyatos1997}, on the other hand, the number of measurements scales as $\mathcal{O}\left(\delta^{-2}\right)$.
In the particular example, analysed numerically here and presented in Fig. (\ref{Simulation1}), we find that the parameters of the phase-damping channel can be estimated with an error of $0.1$, using  $2,500$ measurements per correlation, and $22,500$ measurements in total. This should then be compared with the standard estimation method, that by the scaling argument above is expected to be more efficient. For the same channel and up to the same error, we find that in total only $~1,000$ measurements are required.

\section*{Discussion}

In the present work we have proposed a method for implementing process tomography which is based on the information encoded in temporal correlations. We have shown that such correlations, embodied in the temporal correlations matrix,
provide a general state-independent method to reconstruct the dynamical evolution in terms of Kraus operators.
The complexity of our method grows exponentially with the number of particles in the discrete case (as in the usual approach), and quadratically
with the number of particles in the Gaussian channel case.

Compared with the standard method, our proposed approach is clearly less efficient.
Nevertheless, since it {\em does not} require state preparation, that is essential for ordinary process tomography,
it is applicable in a wider class of cases.
This includes systems in some general mixed states and in particular in thermal states.
In this respect, we hope that the present method could be  beneficial
in various experimental systems, such as NEMS \cite{LaHaye2009,Harvey2010} and mesoscopic systems.

In addition, for systems where the time scales are very short, and the coupling constant between the system and the measuring device is small, it would be difficult to implement strong measurements. Since our method utilizes weak measurement, we believe our method would be suitable for such systems as well.
Weak measurements have been recently realized in various systems \cite{Schuster2005,Laloy2010}.

To further improve the efficiency of our approach, it would be necessary to develop methods to reduce
the error discussed in section 2.
A significant part of this error is systematic,  and may be further reduced by using either analytical or perturbative methods.
For example, when dealing with qubit dynamics and the evolution is known to be unitary, one can calculate $f\left(B_{i}\left(t_{1}\right),B_{j}\left(t_{2}\right)\right)$ and the higher order corrections in terms of the measured correlation and thus eliminate the systematic error. For general, non-unitary, dynamics we believe that  similar results can achieved perturbatively.
Finally,  it would be interesting to investigate whether the present method can be combined with existing efficient methods for selective estimation of dynamics  \cite{Bendersky2008,Schmiegelow2011}.

\section*{Acknowledgments}

The authors would like to thank A.~Botero and R.~Lifshitz for helpful discussions. BR acknowledges
the support of the Israel Science Foundation, the German-Israeli Foundation,
and the European Commission (PICC).

\section*{References}

\bibliographystyle{unsrt}

\begin{thebibliography}{10}


\bibitem{Poyatos1997}
J.F.~Poyatos, J.I.~Cirac, and P.~Zoller.
\newblock Complete Characterization of a Quantum Process: The Two-Bit Quantum Gate.
\newblock {\em Phys. Rev. Lett.}, 78(2):390--393, January 1997.

\bibitem{Emerson2007}
J.~Emerson, M.~Silva, O.~Moussa, C.~Ryan, M.~Laforest, J.Baugh, D.G.Cory, and R.~Laflamme.
\newblock Symmetrized Characterization of Noisy Quantum Processes
\newblock {\em Science}, 317(5846):1893--1896, September 2007.

\bibitem{Bendersky2008}
A.~Bendersky, F.~Pastawski, and J.P.~Paz.
\newblock Selective and Efficient Estimation of Parameters for Quantum Process Tomography.
\newblock {\em Phys. Rev. Lett.}, 100(19):190403 , May 2008.

\bibitem{Schmiegelow2011}
C.T.~Schmiegelow, A.~Bendersky, M.A.~Larotonda, and J.P.~Paz.
\newblock Selective and Efficient Quantum Process Tomography without Ancilla.
\newblock {\em Phys. Rev. Lett.}, 107(10):100502, September 2011.

\bibitem{Emerson2005}
J.~Emerson, R.~Alicki, and K.~\.{Z}yczkowski.
\newblock Scalable noise estimation with random unitary operators.
\newblock {\em J. Opt. B: Quantum Semiclass. Opt.}, 7(10):S347--S352, September 2005.

\bibitem{OperatorNorm}
We use the well known operator norm defined by $\left\Vert A\right\Vert =\underset{v\in V}{\text{max}}\left\{ \frac{\left\Vert Av\right\Vert }{\left\Vert v\right\Vert }\right\}$. For Hermitian operators this norm is simply the largest eigenvalue in absolute value.

\bibitem{Mitchison2007}
G.~Mitchison, R.~Jozsa, and S.~Popescu.
\newblock Sequential weak measurement.
\newblock {\em Phys. Rev. A}, 76(6):062105, December 2007.

\bibitem{Aharonov1988}
Y.~Aharonov, D.Z.~Albert and L.~Vaidman.
\newblock How the result of a measurement of a component of the spin of a spin-1/2 particle can turn out to be 100.
\newblock {\em Phys. Rev. Lett.}, 60(141):1351--1354, April 1988.

\bibitem{Aharonov2005}
Y.~Aharonov and D.~Rohrlich.
\newblock Quantum Paradoxes: Quantum Theory for the Perplexed.
\newblock {\em WILEY-VCH}, 2005

\bibitem{Lidar2001}
D.A.~Lidar, Z.~Bihary, and K.B.~Whaley
\newblock From completely positive maps to the quantum Markovian semigroup master equation.
\newblock {\em Chemical Physics}, 268:35--53, 2001.

\bibitem{Eisert2005}
J.~Eisert and M.M.~Wolf.
\newblock Gaussian quantum channels.
\newblock e-print arXiv:quant-ph/0505151v1.

\bibitem{Holevo2012}
A.S.~Holevo and V.~Giovannetti.
\newblock Quantum channels and their entropic characteristics.
\newblock e-print arXiv:1202.6480v1.

\bibitem{Williamson1936}
J.~Williamson
\newblock On the algebraic problem conceming the normal forms of linear dynamical systems.
\newblock {\em American Journal of Mathematics}, 58:141--163, 1936.

\bibitem{Simon1987}
R.~Simon, E.C.G.~Sudarshan and N.~Mukunda.
\newblock Gaussian-Wigner distributions in quantum mechanics and optics.
\newblock {\em Phys. Rev. A}, 36(8):3868--3880, October 1987.

\bibitem{Holevo2011}
A.S.~Holevo.
\newblock Probabilistic and Statistical Aspects of Quantum Theory.
\newblock {\em Springer}, 2011.

\bibitem{LaHaye2009}
M.D.~LaHaye, J.~Suh, P.M.~Echternach, K.C.~Schwab, and M.L.~Roukes.
\newblock Nanomechanical measurements of a superconducting qubit
\newblock {\em Nature}, 459:960--964, June 2009.

\bibitem{Harvey2010}
T.J.~Harvey, D.A.~Rodrigues, and A.D.~Armour.
\newblock Spectral properties of a resonator driven by a superconducting single-electron transistor.
\newblock {\em Phys. Rev. B}, 81(10):104514, March 2010.

\bibitem{Schuster2005}
D.I.~Schuster, A.~Wallraff, A.~Blais, L.~Frunzio, R.-S.~Huang, J.~Majer, S.M.~Girvin, and R.J.~Schoelkopf.
\newblock ac Stark Shift and Dephasing of a Superconducting Qubit Strongly Coupled to a Cavity Field.
\newblock {\em Phys. Rev. Lett.}, 94(12):123602 , March 2005.

\bibitem{Laloy2010}
A.~Palacios-Laloy, F.~Mallet, F.~Nguyen, P.~Bertet, D.~Vion, D.~Esteve, and A.N.~Korotkov.
\newblock Experimental violation of a Bell\text{'}s inequality in time with weak measurement.
\newblock {\em Nature Physics}, 6:442--447, April 2010.

\end{thebibliography}

\section*{Appendix I: Proof that the covariance matrix is singular iff the state is singular}

Recall that the basis of operators is defined such that $B_{0}=\mathbb{1}_{D\times D}$. First we prove that if
the state $\rho$ is singular then the covariance matrix $\sigma$ is singular: without loss of generality
a singular state can be written in the form
\begin{equation}
\rho=\left(\begin{array}{cc}
\rho_{D-1\times D-1} & 0\\
0 & 0
\end{array}\right)\,.
\end{equation}
We choose $\left\{ B_{i}\right\} $ such that
\begin{equation}
B_{1}=\left(\begin{array}{cc}
0_{D-1\times D-1} & 0\\
0 & 1
\end{array}\right)\,,
\end{equation}
hence $\rho B_{1}=B_{1}\rho=0$, and so
\begin{eqnarray}
\sigma_{1i} & = & \mbox{Tr}\left(\rho\left\{ B_{1},B_{i}\right\} \right)-2\mbox{Tr}\left(\rho B_{1}\right)\mbox{Tr}\left(\rho B_{i}\right)\nonumber \\
 & = & \mbox{Tr}\left(\rho B_{1}B_{i}\right)+\mbox{Tr}\left(B_{1}\rho B_{i}\right)-2\mbox{Tr}\left(\rho B_{1}\right)\mbox{Tr}\left(\rho B_{i}\right)\nonumber \\
 & = & 0\,.
\end{eqnarray}
Since $\sigma_{1i}=\sigma_{i1}=0$ for every $i$, $\sigma$ is singular.

Now we prove that if $\sigma$ is singular then $\rho$ is singular:
if $\sigma$ is singular there must exist a certain linear combination $\alpha_{i} B_{i}$ (which we
refer to as $B_{1}$) whose covariance with every member of $\left\{ B_{i}\right\} $
is zero, and in particular it's variance is zero. Therefore $\left.B_{1}\right|_{\text{range}\left(\rho\right)}=c\!-\!\text{number}$. Since $B_1$ cannot be proportional to the unity $\text{range}\left(\rho\right)<D$,
and thus $\rho$ is singular.

\section*{Appendix II: Correlation measurements using single pointer measurements}

We wish to describe a method for measuring the correlation $\left\{ B_{i}\left(t_{1}\right),B_{j}\left(t_{2}\right)\right\}$ using single pointer measurements.
The interaction between the system and the pointer is of the form:
\begin{eqnarray}
U & = & e^{i\epsilon B_{i}\left(t_{1}\right)\sigma_{2}}e^{i\epsilon B_{j}\left(t_{2}\right)\sigma_{1}},
\end{eqnarray}
where $\epsilon\ll1$. Expanding to second order in $\epsilon$ we
obtain
\begin{equation}
U\cong1-\epsilon^{2}\left(B_{j}^{2}\left(t_{2}\right)+B_{i}^{2}\left(t_{1}\right)\right)+i\epsilon\left(B_{j}\left(t_{2}\right)\sigma_{1}+B_{i}\left(t_{1}\right)\sigma_{2}\right)-\epsilon^{2}B_{i}\left(t_{1}\right)B_{j}\left(t_{2}\right)\sigma_{2}\sigma_{1}.
\end{equation}
Next, we choose
\begin{eqnarray}
\sigma_{1} & = & \frac{1}{\sqrt{2}}\left(\sigma_{z}-\sigma_{x}\right),\\
\sigma_{2} & = & \frac{1}{\sqrt{2}}\left(\sigma_{z}+\sigma_{x}\right),
\end{eqnarray}
hence
\begin{multline}
U\left|\psi_{\text{a,sys}},\uparrow_{x}\right\rangle  = \left(\frac{1}{\sqrt{2}}-\frac{\epsilon^{2}}{\sqrt{2}}\left(B_{j}^{2}\left(t_{2}\right)+B_{i}^{2}\left(t_{1}\right)\right)+i\epsilon B_{i}\left(t_{1}\right)+\frac{1}{\sqrt{2}}\epsilon^{2}B_{i}\left(t_{1}\right)B_{j}\left(t_{2}\right)\right)\left|\psi_{\text{a,sys}},\uparrow_{z}\right\rangle \\
 +\left(\frac{1}{\sqrt{2}}-\frac{\epsilon^{2}}{\sqrt{2}}\left(B_{j}^{2}\left(t_{2}\right)+B_{i}^{2}\left(t_{1}\right)\right)-i\epsilon B_{j}\left(t_{2}\right)-\frac{1}{\sqrt{2}}\epsilon^{2}B_{i}\left(t_{1}\right)B_{j}\left(t_{2}\right)\right)\left|\psi_{\text{a,sys}},\downarrow_{z}\right\rangle .
\end{multline}
Therefore, assuming the initial state is given by $\rho_{\text{in}}=\underset{\text{a}}{\sum}p_{\text{a}}\left|\psi_{\text{a,sys}},\uparrow_{x}\right\rangle \left\langle \psi_{\text{a,sys}},\uparrow_{x}\right|$
the pointers expectation value is
\begin{multline}
E_{1}\left(\sigma_{z}\right) = p_{a}\left\langle \psi_{a}\right|\left(\frac{1}{2}-\frac{\epsilon^{2}}{4}\left(B_{j}^{2}\left(t_{2}\right)+B_{i}^{2}\left(t_{1}\right)\right)+\frac{1}{2}\epsilon^{2}\left\{ B_{i}\left(t_{1}\right),B_{j}\left(t_{2}\right)\right\} +\epsilon^{2}B_{i}^{2}\left(t_{1}\right)\right)\left|\psi_{a}\right\rangle \\
 -p_{a}\left\langle \psi_{a}\right|\left(\frac{1}{2}-\frac{\epsilon^{2}}{4}\left(B_{j}^{2}\left(t_{2}\right)+B_{i}^{2}\left(t_{1}\right)\right)-\frac{1}{2}\epsilon^{2}\left\{ B_{i}\left(t_{1}\right),B_{j}\left(t_{2}\right)\right\} +\epsilon^{2}B_{j}^{2}\left(t_{2}\right)\right)\left|\psi_{a}\right\rangle \\
 = \epsilon^{2}\text{Tr}\left(\rho_{\text{in}}\left\{ B_{i}\left(t_{1}\right),B_{j}\left(t_{2}\right)\right\} \right)+\epsilon^{2}\left(\text{Tr}\left(\rho_{\text{in}}B_{i}^{2}\left(t_{1}\right)\right)-\text{Tr}\left(\rho_{\text{in}}B_{j}^{2}\left(t_{2}\right)\right)\right).
\end{multline}
Now, following the same method with $\sigma_{1}=\frac{1}{\sqrt{2}}\left(\sigma_{z}+\sigma_{x}\right)$
and $\sigma_{2}=\frac{1}{\sqrt{2}}\left(-\sigma_{z}+\sigma_{x}\right)$
we obtain
\begin{equation}
E_{2}\left(\sigma_{z}\right)=\epsilon^{2}\text{Tr}\left(\rho_{\text{in}}\left\{ B_{i}\left(t_{1}\right),B_{j}\left(t_{2}\right)\right\} \right)+\epsilon^{2}\left(-\text{Tr}\left(\rho_{\text{in}}B_{i}^{2}\left(t_{1}\right)\right)+\text{Tr}\left(\rho_{\text{in}}B_{j}^{2}\left(t_{2}\right)\right)\right).
\end{equation}
Summation of these two results yields
\begin{equation}
\frac{1}{2}\left(E_{1}\left(\sigma_{z}\right)+E_{2}\left(\sigma_{z}\right)\right)=\epsilon^{2}\text{Tr}\left(\rho_{\text{in}}\left\{ B_{i}\left(t_{1}\right),B_{j}\left(t_{2}\right)\right\} \right)+\mathcal{O}\!\left(\!\epsilon^{4}\!\right),
\end{equation}
thus it is possible to determine the expectation value of the correlation
$\left\{ B_{i}\left(t_{1}\right),B_{j}\left(t_{2}\right)\right\}$ using
two \emph{one pointer} weak measurements.

\section*{Appendix III: Derivation of Eqs. (\ref{pointers expectation value}-\ref{pointers variance}) and the bound for $\left|f\!\left(B_{i}\!\left(t_{1}\right)\!,\!B_{j}\!\left(t_{2}\right)\right)\right|$}

The derivation of Eqs. (\ref{pointers expectation value}-\ref{pointers variance}) from Eq. (\ref{rho final}) was made straight forward using series expansion in $\epsilon$. This calculation yields
\begin{eqnarray}
\left|f\!\left(B_{i}\!\left(t_{1}\right)\!,\!B_{j}\!\left(t_{2}\right)\right)\right| & = & \left|\frac{1}{48}\text{Tr}\left(\rho_{\text{in}}\left(4\left\{ B_{i}^{3}\left(t_{1}\right),B_{j}\left(t_{2}\right)\right\} +\left\{ B_{i}\left(t_{1}\right),B_{j}^{3}\left(t_{2}\right)\right\} \right)\right)\right.\nonumber \\
 &  & \left.+\frac{1}{16}\text{Tr}\left(\rho_{\text{in}}B_{j}\left(t_{2}\right)\left\{ B_{i}\left(t_{1}\right),B_{j}\left(t_{2}\right)\right\} B_{j}\left(t_{2}\right)\right)\right|\nonumber \\
 & \leq & \frac{1}{3}\left|\max\left\{ \text{Tr}\left(\rho_{\text{in}}B_{k}\left(t'\right)B_{l}\left(t''\right)B_{m}\left(t'''\right)B_{n}\left(t''''\right)\right)\right\}\right|\nonumber \\
 & \leq & \frac{1}{3}\max\left\{ \left\Vert B_{k}\left(t'\right)B_{l}\left(t''\right)B_{m}\left(t'''\right)B_{n}\left(t''''\right)\right\Vert \right\}
\end{eqnarray}
where $k,l,m,n\in\left\{ i,j\right\} $ and $t',t'',t''',t''''\in\left\{ t_{1},t_{2}\right\} $.
Now we choose all operators in time $t_{1}$ to be of the same norm, $\left\Vert B_{k}\left(t_{1}\right)\right\Vert \equiv\left\Vert B\right\Vert$,
and since Quantum evolution guarantees $\left\Vert B_{k}\left(t_{2}\right)\right\Vert \leq\left\Vert B_{k}\left(t_{1}\right)\right\Vert $
for every $k$, we obtain
\begin{eqnarray}
\left|f\left(B_{i}\left(t_{1}\right),B_{j}\left(t_{2}\right)\right)\right| & \leq & \frac{1}{3}\left\Vert B^{4}\right\Vert\nonumber \\
 & \leq & \frac{1}{3}\left\Vert B\right\Vert ^{4}
\end{eqnarray}

\end{document}